\documentclass[aps,showpacs,amsfonts,amsmath, amssymb,twocolumn,floatfix, superscriptaddress]{revtex4}
\usepackage[final]{graphicx}
\usepackage{xcolor}
\usepackage[colorlinks=true, linkcolor=blue, citecolor=blue, urlcolor=blue]{hyperref}
\usepackage{verbatim}
\usepackage{csquotes}
\usepackage{braket}
\usepackage{mathtools}

\newcommand{\vct}[1]{\mathbf{#1}}

\renewcommand\Re{\operatorname{Re}}
\renewcommand\Im{\operatorname{Im}}
\newcommand\Tr{{\rm Tr}}

\newcommand{\be}{\begin{equation}}
\newcommand{\ee}{\end{equation}}
\allowdisplaybreaks

\DeclareSymbolFont{bbgreek}{U}{bbold}{m}{n}
\DeclareMathSymbol{\bbmu}{\mathbb}{bbgreek}{'26}
\DeclareMathSymbol{\bbeps}{\mathbb}{bbgreek}{'17}

\begin{document}


\title{Heat radiation and transfer in confinement}

\author{Kiryl Asheichyk}
\email[]{asheichyk@is.mpg.de}
\affiliation{4th Institute for Theoretical Physics, Universit\"at Stuttgart, Pfaffenwaldring 57, 70569 Stuttgart, Germany}
\affiliation{Max Planck Institute for Intelligent Systems, Heisenbergstrasse 3, 70569 Stuttgart, Germany}   
\author{Matthias Kr\"uger}
\email[]{matthias.kruger@uni-goettingen.de}
\affiliation{Institute for Theoretical Physics, Georg-August-Universit\"at G\"ottingen, 37073 G\"ottingen, Germany}

\begin{abstract} 
Near-field heat radiation and transfer are rich in various exciting effects, in particular, regarding the amplification due to the geometrical configuration of the system. In this paper, we study heat exchange in situations where the objects are confined by additional objects so that the dimensionality of heat flow is reduced. In particular, we compute the heat transfer for spherical point particles placed between two parallel plates. The presence of the plates can enhance or reduce the transfer compared to the free case and provides a slower power-law decay for large distance. We also compute the heat radiation of a sphere placed inside a spherical cavity, finding that it can be larger or smaller compared to the radiation of a free sphere. This radiation shows strong resonances as a function of the cavity's size. For example, the cooling rate of a nanosphere placed in a cavity varies by a factor of $10^5$ between cavity radii $ 2 \ \mu {\rm m} $  and $ 5 \ \mu {\rm m} $. 
\end{abstract}


\bibliographystyle{plain}

\maketitle



\section{Introduction}
\label{sec:Introduction}
Development of fluctuational electrodynamics~\cite{Rytov1957, Rytov1989} and improvement of theoretical and numerical techniques for electromagnetic scattering theory allowed to explore a vast amount of effects for heat radiation (HR) and radiative heat transfer (HT) for complex 
objects~\cite{Tsang2004, Bohren2004, Tsang2001, Rodriguez2011, McCauley2012, Rodriguez2013, Polimeridis2015}. Moreover, a significant effort has been made to generalize the theory and develop formalisms which can be used to compute HR, 
HT, and nonequilibrium Casimir forces in arbitrary many-body systems~\cite{Rahi2009, Kruger2011, Messina2011_1, Messina2011_2, Messina2014, Kruger2012, Muller2017, Bimonte2017}. It is, however, 
challenging to study complex geometrical configurations as they require solutions to non-trivial boundary conditions problems and may also require long computational times despite the existence of sophisticated numerical methods~\cite{Rodriguez2011, McCauley2012, Rodriguez2013, Polimeridis2015}.

Typically, to investigate many-body effects, one applies certain simplifications for the system allowing to predict realistic results without significant loss of generality. One of the most popular 
simplifications is the point particle limit where HR and HT are computed for pointlike particles~\cite{Ben-Abdallah2011, Dong2017, Asheichyk2017, Dong2018, Messina2018}. In this case, it has been shown that the presence of
an additional object can have a large effect on the heat exchange between the particles, including strong enhancements of the HT compared to the vacuum case~\cite{Ben-Abdallah2011, Dong2017, Asheichyk2017, Dong2018, Messina2018}. Also, in this case, the boundary 
conditions can be described by the Green's function of the objects surrounding the particles~\cite{Ben-Abdallah2011, Asheichyk2017}.

Up to now, mostly open systems have been considered, such that  heat can flow in all directions.  In this work, we study HR and HT in closed systems, based on the formalisms developed in Refs.~\cite{Kruger2012, Muller2017,Asheichyk2017}. Specifically, we discuss two paradigmatic 
closed systems, namely the HT between two point particles confined by two parallel plates and the HR of a sphere of arbitrary size enclosed by a spherical cavity. We observe that confinement has a large effect on HR and HT and can increase or decrease these quantities by several 
orders of magnitude compared to isolated objects. Moreover, in the case of HR inside a spherical cavity, we observe a strongly nonmonotonic behavior of a sphere's heat emission indicating that the 
cavity acts as a resonator. The cavity can thus be used as an insulated bag as well as a cooler or heater for nanoparticles, with orders of magnitude faster cooling or heating rates compared to the vacuum case.

The paper is structured as follows. In Sec.~\ref{sec:HT_confinement_1D}, we study the HT between two nanoparticles placed inside a two-plates cavity. Section~\ref{sec:HT_confinement_3D}  investigates the HR of a sphere enclosed by a spherical cavity. The paper is closed with a 
summary and discussion in Sec.~\ref{sec:Conclusion}.

\section{Removing one dimension: a cavity between two parallel plates}
\label{sec:HT_confinement_1D} 
In this section, we discuss the HT in a cavity made by two parallel plates. The discussion is based on fluctuational electrodynamics and scattering theory. For details on the formalism employed, we refer the reader to Refs.~\cite{Kruger2012, Muller2017, Asheichyk2017}. 

\begin{figure}[!t]
\includegraphics[width=0.8\linewidth]{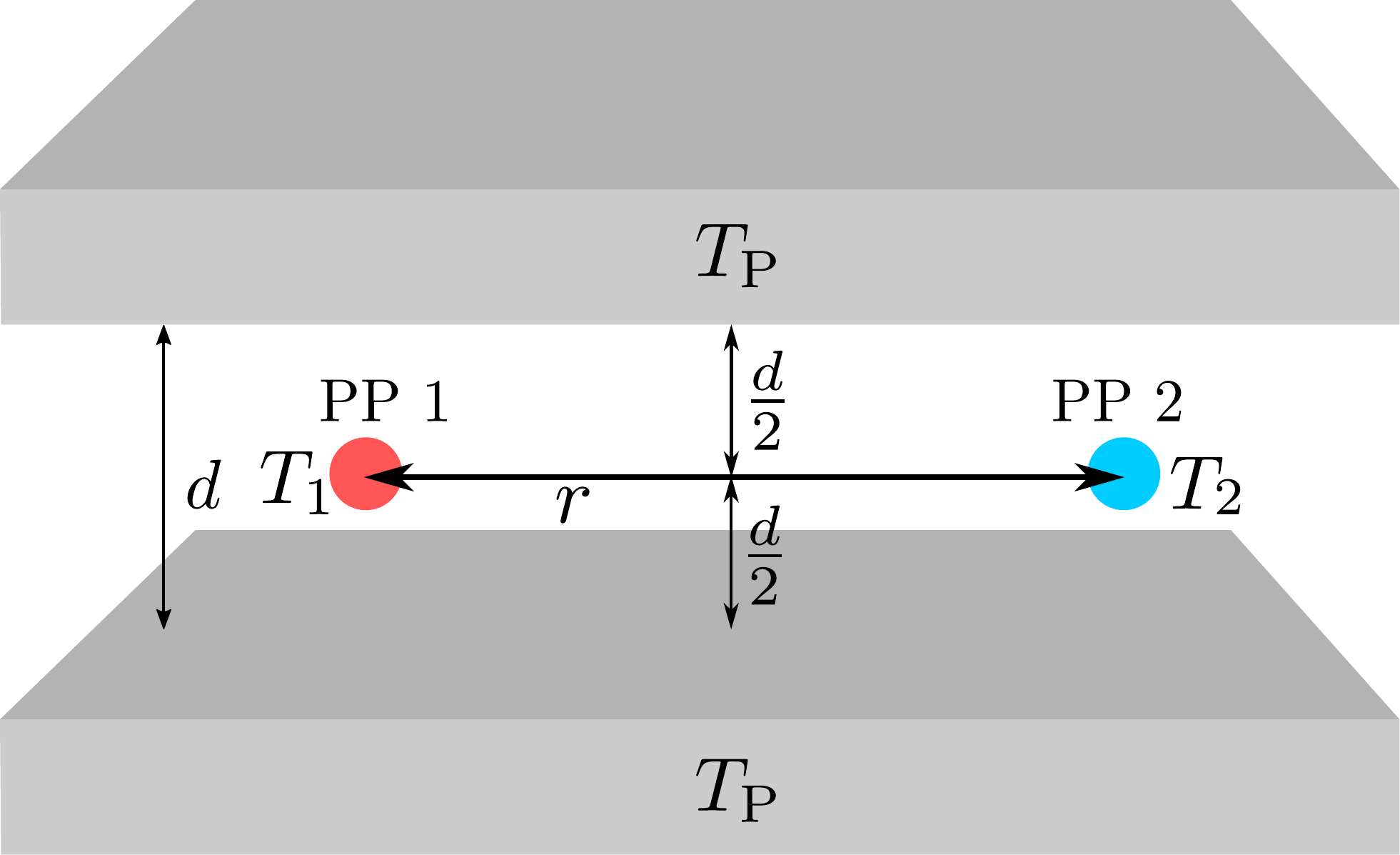}
\caption{\label{fig:config_plates}
Two identical semi-infinite parallel plates exemplifying systems that remove one dimension. We aim to compute the heat transfer from point particle 1 (PP~1) to point particle 2 (PP~2) in this system. 
$ T_1 $, $ T_2 $, and $ T_{\rm P} $ are temperatures of the particles and the plates, respectively. For the quantity we compute, only $ T_1 $ is relevant.}
\end{figure}
The simplest system that effectively removes one dimension for heat flow is a two-plates cavity as depicted in Fig.~\ref{fig:config_plates}.  Our goal is to compute the HT from point particle 1 (PP~1) to point particle 2 (PP~2) in this system. We note that the point particle limit is valid if the radius of
each particle is small compared to any other length scale in the system related to the particles~\cite{Asheichyk2017}, including thermal wavelength, skin penetration depth of each particle, the distance
between the particles $ r $, and the distance from each particle to a plate $ \frac{d}{2} $. In this limit,  the multiple scatterings from the particles can be neglected and the particles can be modeled 
by the electrical dipole polarizability~\cite{Asheichyk2017}. The HT reads as (quoted from Ref.~\cite{Asheichyk2017})
\begin{align}
\notag H_{1pp}^{(2pp)} = & \ \frac{32\pi\hbar}{c^4} \int_0^\infty d\omega \frac{\omega^5}{e^{\frac{\hbar\omega}{k_{\rm B}T_1}}-1}\Im(\alpha_1)\\ 
&\times \Im(\alpha_2)\sum_{ij}|G_{ij}(\vct{r}_2, \vct{r}_1)|^2.
\label{eq:HT_PPs}
\end{align}
Here, $ G_{ij}(\vct{r}_2, \vct{r}_1) $ is a matrix element of the Green's function (GF) of the plates $ \mathbb{G} $, where $ \vct{r}_1 $ and $ \vct{r}_2 $ are the coordinates of the particles. For the 
configuration depicted in Fig.~\ref{fig:config_plates}, the GF is given by Eq.~\eqref{eq:GF2PlatesFinal} in Appendix~\ref{app:GF_plates}. $ \alpha_i $ is a particle's polarizability given by
\begin{equation}
\alpha_i = \frac{\varepsilon_i-1}{\varepsilon_i+2}R_i^3
\label{eq:polarizability},
\end{equation}
where $ \varepsilon_i $ is the dielectric function and $ R_i $ is the radius of $ i $th particle, respectively. $ T_1 $ is the temperature of the first particle, $ c $ is the speed of light in vacuum, 
$ \hbar $ and $ k_{\rm B} $ are Planck's and Boltzmann's constants, respectively. 
To achieve maximum symmetry of the configuration, we place the particles in a plane parallel to the plates and located exactly in the center between them as shown in Fig.~\ref{fig:config_plates}. 

We emphasize that the quantity we compute, $ H_{1pp}^{(2pp)} $, is the HT from PP~1 to PP~2, i.e., the rate of heat emitted by PP~1 and absorbed by PP~2. However, there are, in general, other heat flow 
contributions, e.g., the heat transfer from PP~1 to the plates or from PP~2 to PP~1. For example, the net heat radiated by PP~2, $ H^{(2pp)} $, which is an experimentally accessible quantity, 
includes the heat radiation of PP~2 (the heat transfer from PP~2 to itself) as well as heat transfers from PP~1 to PP~2 and from the plates to PP~2~\cite{Kruger2012}. In 
the case where $ T_2 = T_{\rm P} = 0 $, $ H_{1pp}^{(2pp)} = - H^{(2pp)} $. Since heat transfer 
contributions are independent, the HT from PP~1 to PP~2 is not affected by the presence of other heat flows~\cite{Kruger2012}. Therefore, $ H_{1pp}^{(2pp)} $ depends only on the temperature $ T_1 $ of 
PP~1, and other temperatures in the system, e.g., that of plates, are hence irrelevant for our computations. Similar discussions apply for the system in Fig.~\ref{fig:config_cavity}. For detailed 
discussions regarding different heat flow contributions, we refer the reader to Ref.~\cite{Kruger2012}.

\begin{figure}[!t]
\includegraphics[width=1.0\linewidth]{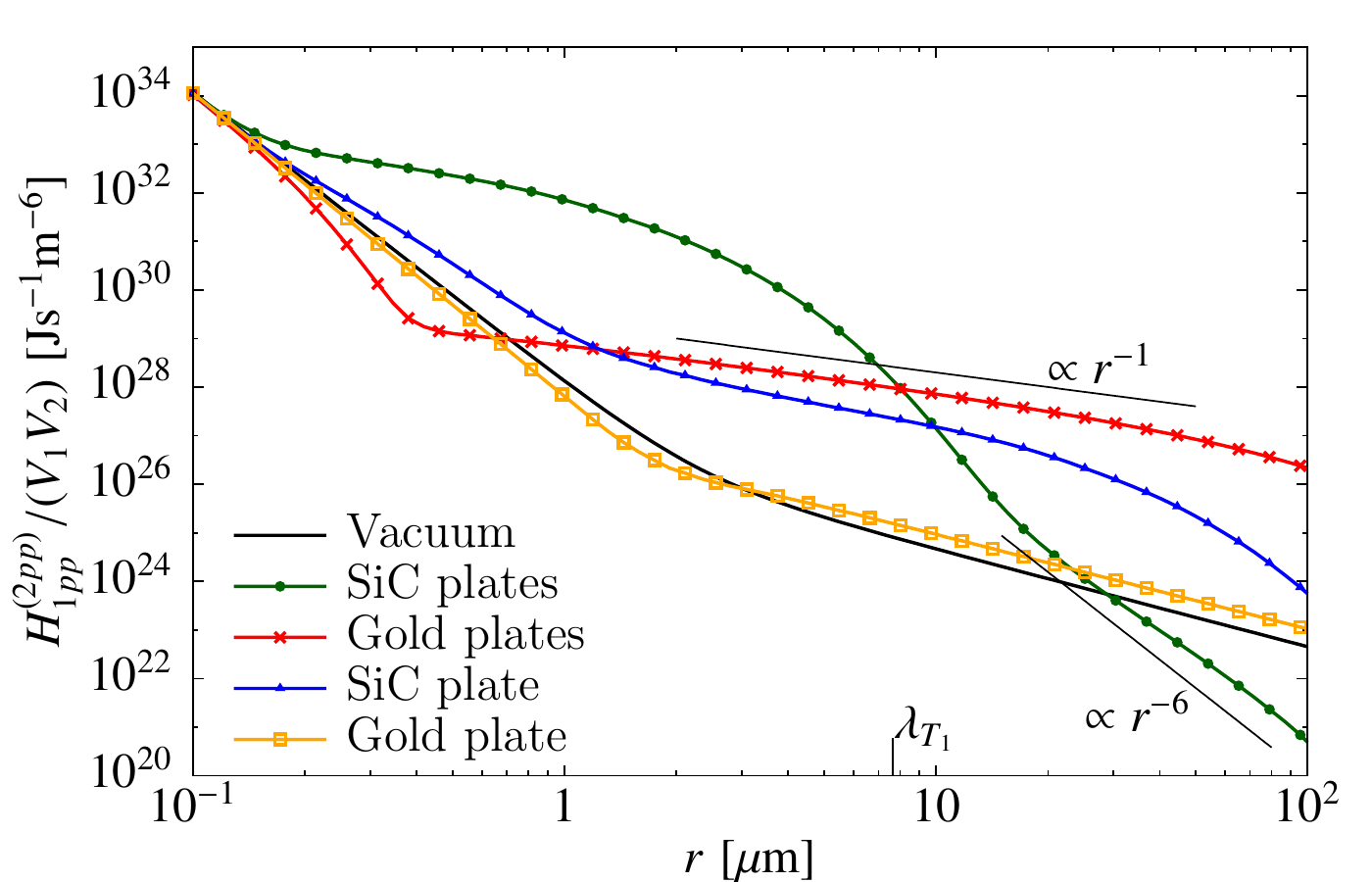}
\caption{\label{fig:plates2_HT}
Normalized (by the volumes of the particles) HT from SiC PP~1 at temperature $ T_1 = 300 \ {\rm K} $ to SiC PP~2 inside a two-plates cavity (see Fig.~\ref{fig:config_plates}) as a function of the 
distance $ r $ between the particles. The distance $ d $ between the plates is fixed at $ d = 2 \times 10^{-1} \ \mu{\rm m} $. The result is compared to the HT in vacuum and in the presence of a single 
plate. Thermal wavelength $ \lambda_{T_1} \approx 7.63 \ \mu{\rm m} $.}
\end{figure}
Figure \ref{fig:plates2_HT} shows specific results for two SiC particles  with $ d = 2 \times 10^{-1} \ \mu{\rm m} $, such that each particle is at the distance $ h = \frac{d}{2} = 10^{-1} \ \mu {\rm m} $ from each plate, as a function of the distance $ r $ between the particles. The temperature of PP~1 is $ T_1 = 300 \ {\rm K} $. We consider the plates to be made of SiC or gold. For SiC, we use the 
following dielectric function~\cite{Spitzer1959}:
\begin{equation}
\varepsilon_{\rm SiC}(\omega) = \varepsilon_\infty\frac{\omega^2-\omega_{\rm LO}^2+i\omega\gamma}{\omega^2-\omega_{\rm TO}^2+i\omega\gamma},
\label{eq:epsilon_SiC}
\end{equation}
where  $ \varepsilon_\infty=6.7 $, $ \omega_{\rm LO}=1.82\times10^{14} \ {\rm rad} \ {\rm s}^{-1} $, $ \omega_{\rm TO}=1.48\times10^{14} \ {\rm rad} \ {\rm s}^{-1} $, 
$ \gamma=8.93\times10^{11} \ {\rm rad} \ {\rm s}^{-1} $. For gold, the Drude model was used,
\begin{equation}
\varepsilon_{\rm Au}(\omega) = 1-\frac{\omega_p^2}{\omega(\omega+i\omega_{\tau})},
\label{eq:epsilon_gold}
\end{equation}
with $ \omega_p = 1.37\times 10^{16} \ {\rm rad} \ {\rm s}^{-1} $ and $ \omega_{\tau} = 4.06\times 10^{13} \ {\rm rad} \ {\rm s}^{-1} $. Since in the point particle limit the HT is proportional to the 
volumes $ V_1 $ and $ V_2 $ of the particles [see Eqs.~\eqref{eq:HT_PPs} and~\eqref{eq:polarizability}], we do not give the particles' sizes explicitly and normalize the curves by their volumes. For our 
configuration, the PP limit applies for $ R_i \lessapprox 10 \ {\rm nm} $. The results are compared to the HT in the presence of a single plate (where all parameters are the same, i.e., we remove one 
plate from the system without changing other parameters) and the HT for the particles in free space. See, e.g., Ref.~\cite{Messina2018} for the GF of a single plate.

For two SiC plates, the HT  is larger than the vacuum HT up to $ r \approx 30 \ \mu{\rm m} $. The enhancement is very large and exceeds a factor of four orders of magnitude (at around $ r = 2 \ \mu{\rm m} $). For $ r \gtrapprox 30 \ \mu{\rm m} $, the HT is smaller than that for isolated particles and decays as $ \sim r^{-6} $.  The ultimate behavior for $ r\to\infty $ remains unknown. In contrast to the two-plates case, the HT in the presence of a single plate, as studied in Refs.~\cite{Dong2018, Messina2018}, shows a lower, but a longer (in distance $ r $) enhancement. The presence of plates is thus very nonadditive, i.e., the transfer with two plates can be remarkably different compared to a single plate, demonstrating the presence of confined modes (distinct from the surface modes present for a single plate \cite{Dong2018, Messina2018}).

This statement is even more true for the case where the plates are made of gold. Here, a single plate has almost no effect, as no surface modes are excited. However, the HT is largely enhanced by two plates for $ r \gtrapprox 1 \ \mu{\rm m} $. For larger $r$, we observe a decay with $ \sim r^{-1} $, which we attribute to energy conservation. In free space, the energy emitted by the first particle distributes over a spherical surface, so that the HT in free space decays as $ r^{-2} $. In the confined situation, the cavity acts as a wave guide, and energy distributes over a circle, leading to the decay with $ r^{-1} $. At $ r \approx 10^2 \ \mu{\rm m} $, this power law is cut off by the mechanism of absorption of waves by the gold surfaces (the wave guide is imperfect). We expect this power law of $ r^{-1} $ to extend to infinity for the case of perfect mirror plates (whose numerical evaluation is, however, nontrivial). 
\begin{figure}[!t]
\includegraphics[width=1.0\linewidth]{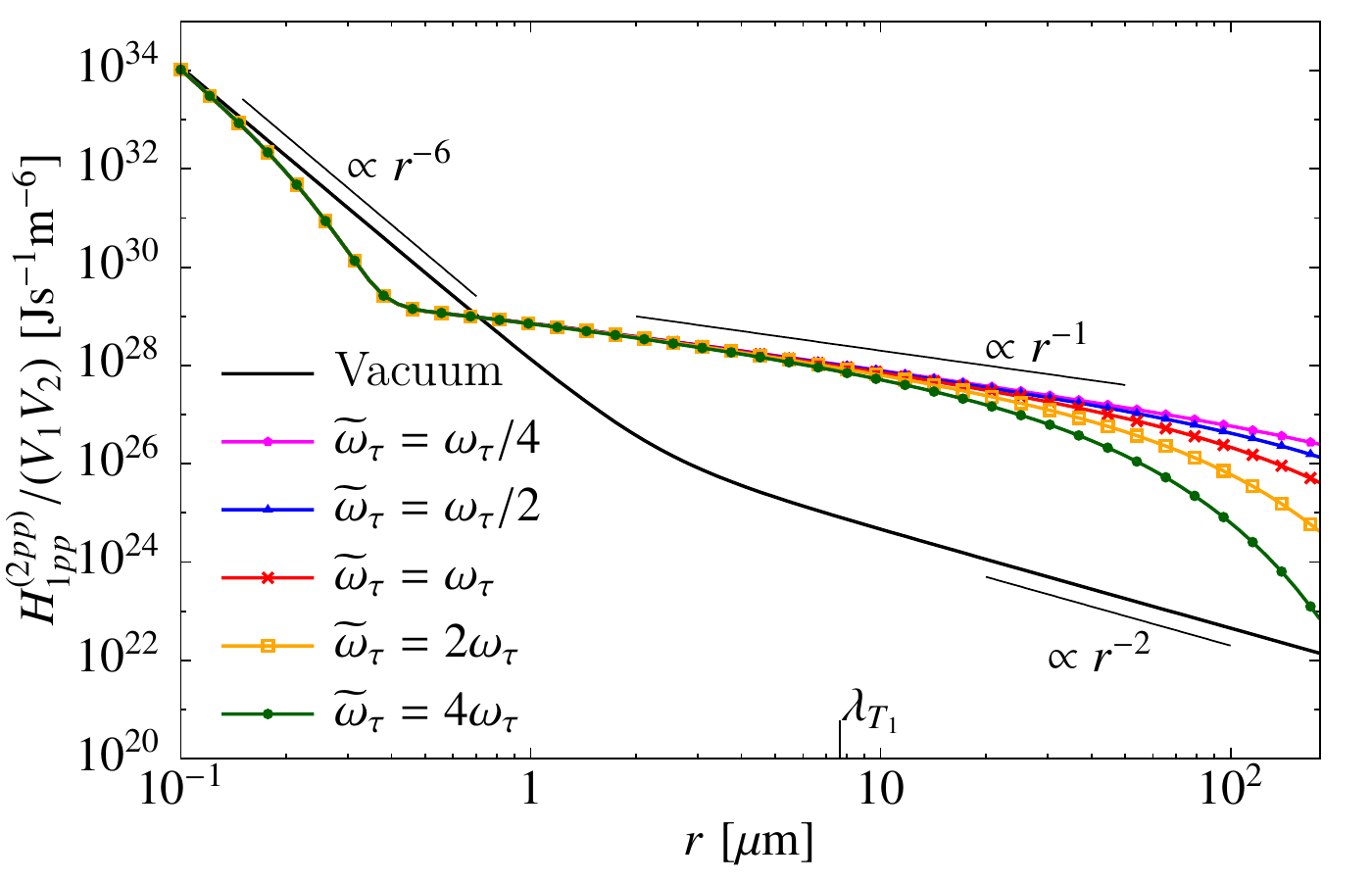}
\caption{\label{fig:diff_damp}
The heat transfer between two point particles inside a two-plates cavity made by metallic plates with different damping rates $\widetilde{\omega}_{\tau}$. The case 
$\widetilde{\omega}_{\tau} = \omega_{\tau}$ corresponds to gold plates. Other parameters are as in Fig.~\ref{fig:plates2_HT}.}
\end{figure}
To underpin this expectation, we varied the damping rate $\omega_{\tau}$ in Eq.~\eqref{eq:epsilon_gold}, replacing it by $\widetilde{\omega}_{\tau}$. The resulting HT is shown in 
Fig.~\ref{fig:diff_damp}. The figure shows that the quality of the two-plates wave guide increases with decrease of $\widetilde{\omega}_{\tau}$, i.e., the $r^{-1}$ power law lasts longer with decrease 
of the damping.

Lastly, we note that the HT decreases monotonically for all the considered cases. This is in contrast to the HT in the presence of a sphere~\cite{Asheichyk2017}. The results of Fig.~\ref{fig:plates2_HT} 
for a single plate are in agreement with Refs.~\cite{Dong2018, Messina2018}.

\section{Removing three dimensions: a spherical cavity}
\label{sec:HT_confinement_3D}
\begin{figure}[!t]
\includegraphics[width=0.6\linewidth]{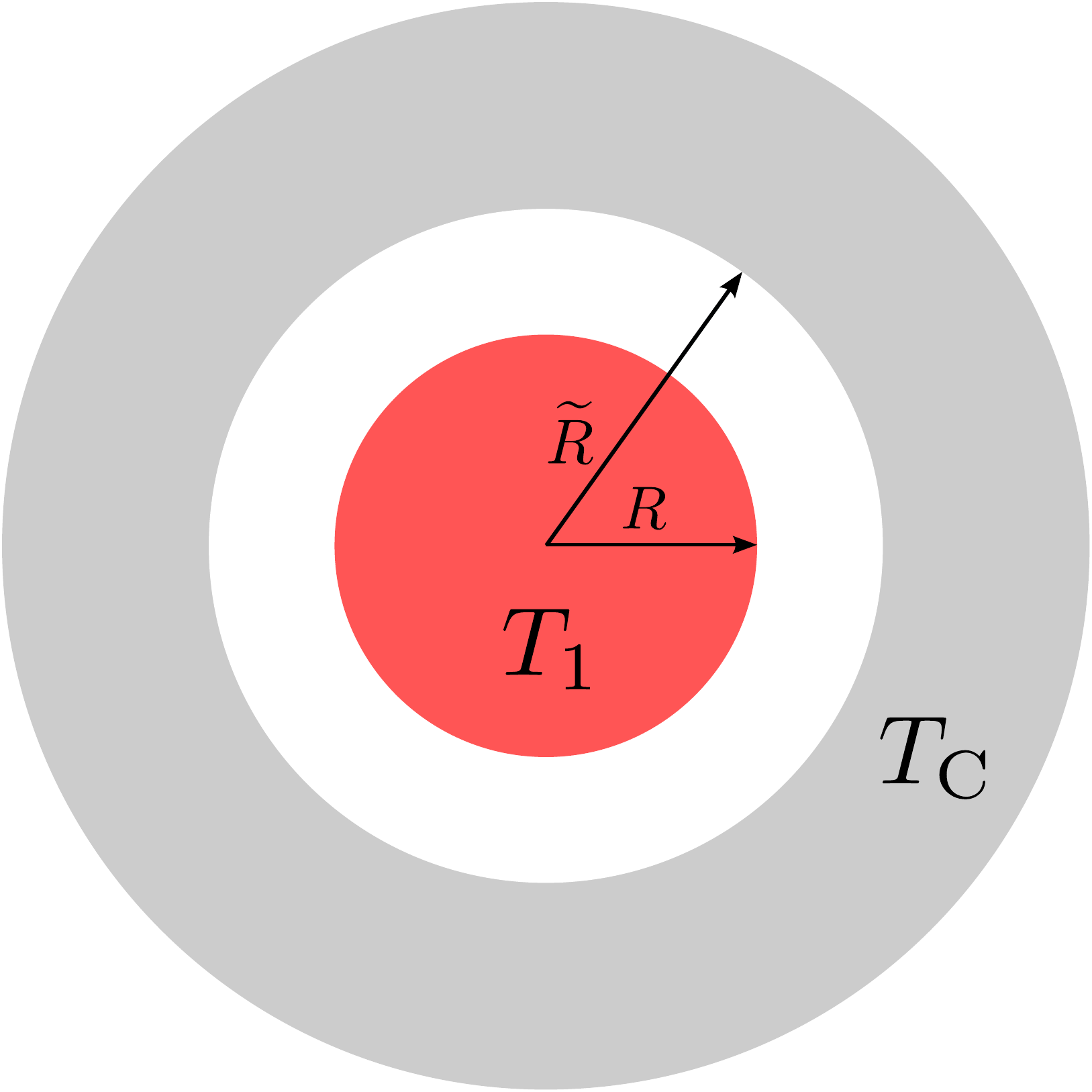}
\caption{\label{fig:config_cavity}
A spherical cavity exemplifying a system closed in all three dimensions. We aim to compute the heat radiation of a sphere placed in the center of the cavity. 
$ T_1 $ and $ T_{\rm C} $ are temperatures of the sphere and the cavity, respectively. For the quantity we compute, only $ T_1 $ is relevant.}
\end{figure}
Energy confinement in all directions can be achieved by placing an object inside a cavity. The most symmetric configuration for such scenario is a 
sphere placed in the center of a spherical cavity (see Fig.~\ref{fig:config_cavity}). The wall of the cavity is assumed to be infinitely extended. The formula for the HR in 
this geometry is derived in Appendix~\ref{app:HR_sphere_cavity_derivation}. Due to the symmetry of the system, this formula takes a particularly simple form (reminiscent of the result for two parallel surfaces) and reads as
\begin{align}
\notag {H_1^{(1)}} = & -\frac{2\hbar}{\pi} \int_0^\infty d\omega \frac{\omega}{e^{\frac{\hbar\omega}{k_BT_1}}-1} \sum_{P=M,N}\\
&\times \sum_{l=1}^{\infty}(2l+1)\frac{\big(\Re \widetilde{\mathcal{T}}_l^P + 1\big)\big(\Re \mathcal{T}_l^P + {\left|\mathcal{T}_l^P\right|}^2\big)}{{\left|1-\widetilde{\mathcal{T}}_{l}^P\mathcal{T}_l^P\right|}^2},
\label{eq:HR_sphere_cavity}
\end{align}
where $ \mathcal{T}_l^P $ are the scattering matrix elements of a sphere of order $ l $ and polarization $ P $ given by Eqs.~(B8) and~(B9) in 
Ref.~\cite{Asheichyk2017}, $ \widetilde{\mathcal{T}}_l^P $ are the scattering matrix elements of a cavity given in Ref.~\cite{Zaheer2010}, and $ T_1 $ is the temperature of a sphere. For consistency with Refs.~\cite{Kruger2012, Muller2017, Asheichyk2017}, the subscript and the 
superscript $ 1 $ on the left-hand side of Eq.~\eqref{eq:HR_sphere_cavity} state the label of a sphere, i.e., a sphere is object~$ 1 $.

\begin{figure}[!t]
\includegraphics[width=1.0\linewidth]{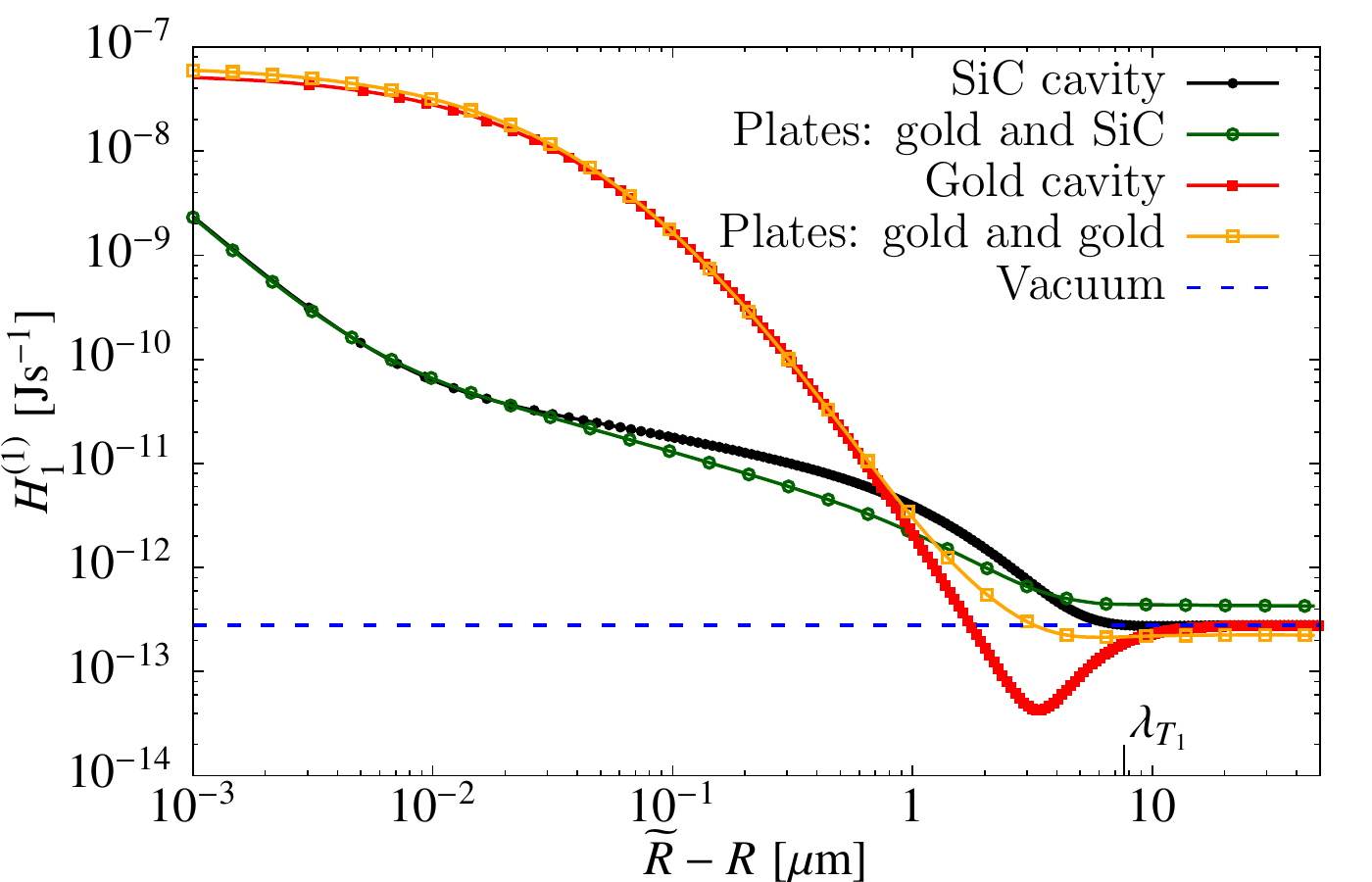}
\caption{\label{fig:CavityHR_gold_sph}
Heat radiation of a gold sphere with radius $ R = 10^{-1} \ \mu{\rm m} $ and temperature $ T_1 = 300 \ {\rm K} $ placed in the center of a spherical cavity with radius $ \widetilde{R} $ as a function of the distance 
$ \widetilde{R} - R $ between the sphere's surface and the cavity wall (made of SiC or gold), see Fig.~\ref{fig:config_cavity}. The result is compared to the heat transfer between two parallel plates at distance $ \widetilde{R} - R $, evaluated for the sphere's surface area. The dashed 
line corresponds to the HR of the sphere in isolation.}
\end{figure}

We emphasize that the quantity we compute, $ H_1^{(1)} $, is the HR of the sphere, i.e., the rate of heat emitted by the sphere and absorbed by it. Therefore, only $ T_1 $ is relevant 
for our computations. To find the net heat radiation (which is minus the net absorption) of the sphere, $ H^{(1)} $, one should also include the heat transfer from the cavity to the sphere, 
$ H_C^{(1)} $, such that the resulting HR is $ H^{(1)} = H_1^{(1)} - H_C^{(1)} $. Due to the symmetry of the system, $ H_C^{(1)} = H_1^{(1)}$ with $ T_1 $ replaced by $ T_C $ in Eq.~\eqref{eq:HR_sphere_cavity}.

As a consistency check, we trivially observe that the result of  Eq.~\eqref{eq:HR_sphere_cavity} vanishes if $ \mathcal{T}_l^P \to 0 $ (the sphere becomes transparent), as required. In the opposite limit of a perfectly reflecting sphere, one has~\cite{Kruger2012}
\begin{equation}
\lim_{|\varepsilon| \to \infty} \Re \mathcal{T}_l^P = - \lim_{|\varepsilon| \to \infty} {\left|\mathcal{T}_l^P\right|}^2,
\label{eq:T_mirror_sphere}
\end{equation}
and therefore
\begin{equation}
\lim_{|\varepsilon| \to \infty} {H_1^{(1)}} = 0,
\label{eq:HR_mirror_sph_cavity}
\end{equation}
as expected as well (a perfectly reflecting sphere does not radiate energy). Equally expected, if there is no cavity, $ \widetilde{\mathcal{T}}_l^P = 0 $, and Eq.~\eqref{eq:HR_sphere_cavity} equals the HR of a sphere in isolation (see e.g. Eq.~(124) in Ref.~\cite{Kruger2012}). 
On the other hand, in the perfect mirror limit for the wall of the cavity, $ \Re \lim_{|\widetilde{\varepsilon}| \to \infty}\widetilde{\mathcal{T}}_{l}^P = -1 $ 
(see Appendix~B in Ref.~\cite{Asheichyk2017}), and hence
\begin{equation}
\lim_{|\widetilde{\varepsilon}| \to \infty} {H_1^{(1)}} = 0.
\label{eq:HR_sph_mirror_cavity}
\end{equation}
In that limit, all energy emitted by the sphere comes back to it with the opposite sign of the energy flow, i.e., the net energy flow is zero. 

Finally, we note that, since the system is completely closed, the HR of the sphere can also be interpreted as the HT from the sphere to the cavity.

\begin{figure}[!t]
\includegraphics[width=1.0\linewidth]{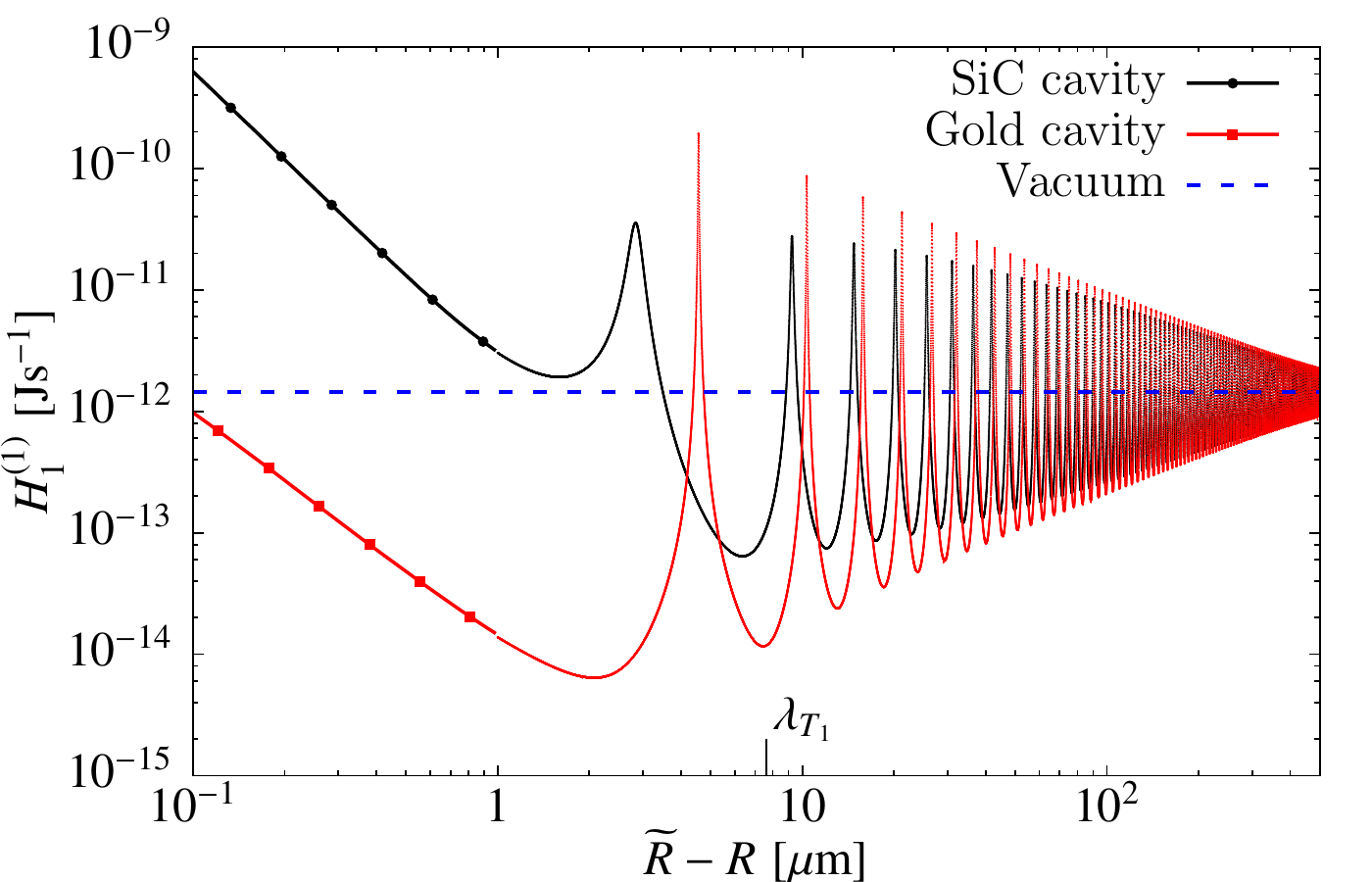}
\caption{\label{fig:CavityHR_SiC_sph}
Heat radiation of a SiC sphere inside a spherical cavity, with all parameters as in Fig.~\ref{fig:CavityHR_gold_sph}.}
\end{figure}

Figure~\ref{fig:CavityHR_gold_sph} demonstrates the case of a gold sphere of radius $ R = 10^{-1} \ \mu{\rm m} $ and temperature $ T_1 = 300 \ {\rm K} $ inside a cavity (SiC or gold) of radius $ \widetilde{R} $, shown as a function  of the distance 
$ \widetilde{R} - R $ between the sphere's surface and the surface of the cavity. When evaluating Eq.~\eqref{eq:HR_sphere_cavity}, the sum over $ l $ is truncated at sufficient order (see Fig.~\ref{fig:CavityHR_gold_sph_conv} below). 
For small $ \widetilde{R} - R $, we observe a large enhancement for both gold and SiC cavity, which can be understood from the HT between two parallel surfaces. If the distance between surfaces of sphere and cavity is small compared to their radii, we may expect that the HR can be expressed via the result for two parallel plates (so called proximity approximation \cite{Kruger2011, Gies2006, Narayanaswamy2008, Otey2011, Sasihithlu2011, Golyk2012}). Indeed, the result for two parallel plates, evaluated for the surface area of the sphere, fits well to Eq.~\eqref{eq:HR_sphere_cavity}, as shown in the graph. 
In the opposite limit, for $ \widetilde{R} - R \gtrapprox \lambda_{T_1}$, we observe the approach of the result for an isolated sphere. In this limit, the radiation reflected back from the cavity to the sphere scatters many times from the cavity wall, so that the cavity has the same effect as a (black-body) environment. This is different for two parallel plates, where the HT between the plates separated by a large distance is still distinct from the emission of a single surface.  
For a SiC cavity, the curve monotonically interpolates between these two limits, and the estimate from two parallel plates is always within $50\ \%$ error. However, for gold, we observe a pronounced minimum in between, where the HT is suppressed by roughly one order of magnitude. This can be understood from the insights around Eq.~\eqref{eq:HR_sph_mirror_cavity}: the reflectivity ($|\widetilde{\varepsilon}|$) of a gold cavity is quite high, so that the HR is suppressed. 

The situation is rather different for a SiC sphere, as presented in Fig.~\ref{fig:CavityHR_SiC_sph}, with all other parameters as in Fig.~\ref{fig:CavityHR_gold_sph}. The curves show pronounced peaks for distances around and beyond the thermal wavelength $\lambda_{T_1}$. While a SiC sphere placed in a gold cavity of $R\approx 2 \ \mu{\rm m}$ emits roughly $5\times 10^{-3}$ times the value of a free sphere, this factor is about $10^2$ in a cavity of $R\approx 5 \ \mu{\rm m}$. The cavity may thus be used to insulate the sphere or to speed up cooling. 
We attribute these peaks to resonances of the cavity, so that they occur if multiply reflected waves add constructively. The emissivity of a SiC sphere is strongly peaked at a wavelength of  $ \lambda_0 \approx 10.75 \ \mu{\rm m} $ , and the HR indeed roughly peaks at half multiples of this value. In contrast to that, a gold sphere has a very broad emissivity at room temperature, so that these resonances are not visible in Fig.~\ref{fig:CavityHR_gold_sph}. 
These resonances also strongly delay the approach of the free sphere for large $ \widetilde{R} - R $. While for a gold sphere in Fig.~\ref{fig:CavityHR_gold_sph}, the radiation equals that of the free sphere for $ R  \gtrapprox 10 \ \mu{\rm m}$, it  takes values of $ R $ in the range of $10^3 \ \mu{\rm m}$ for a SiC sphere to approach that limit. We thus observe nontrivial HR effects for distances of millimeters. The result in Fig.~\ref{fig:CavityHR_SiC_sph} is an example of an electromagnetic resonator in the context of heat radiation. It shows that such a closed system can provide strongly nonmonotonic behavior for HR and HT such that resulting quantities are very sensitive to small changes of the system's parameters. Since, in reality, most systems are closed, such results may be important for a large variety of applications.

\begin{figure}[!t]
\includegraphics[width=1.0\linewidth]{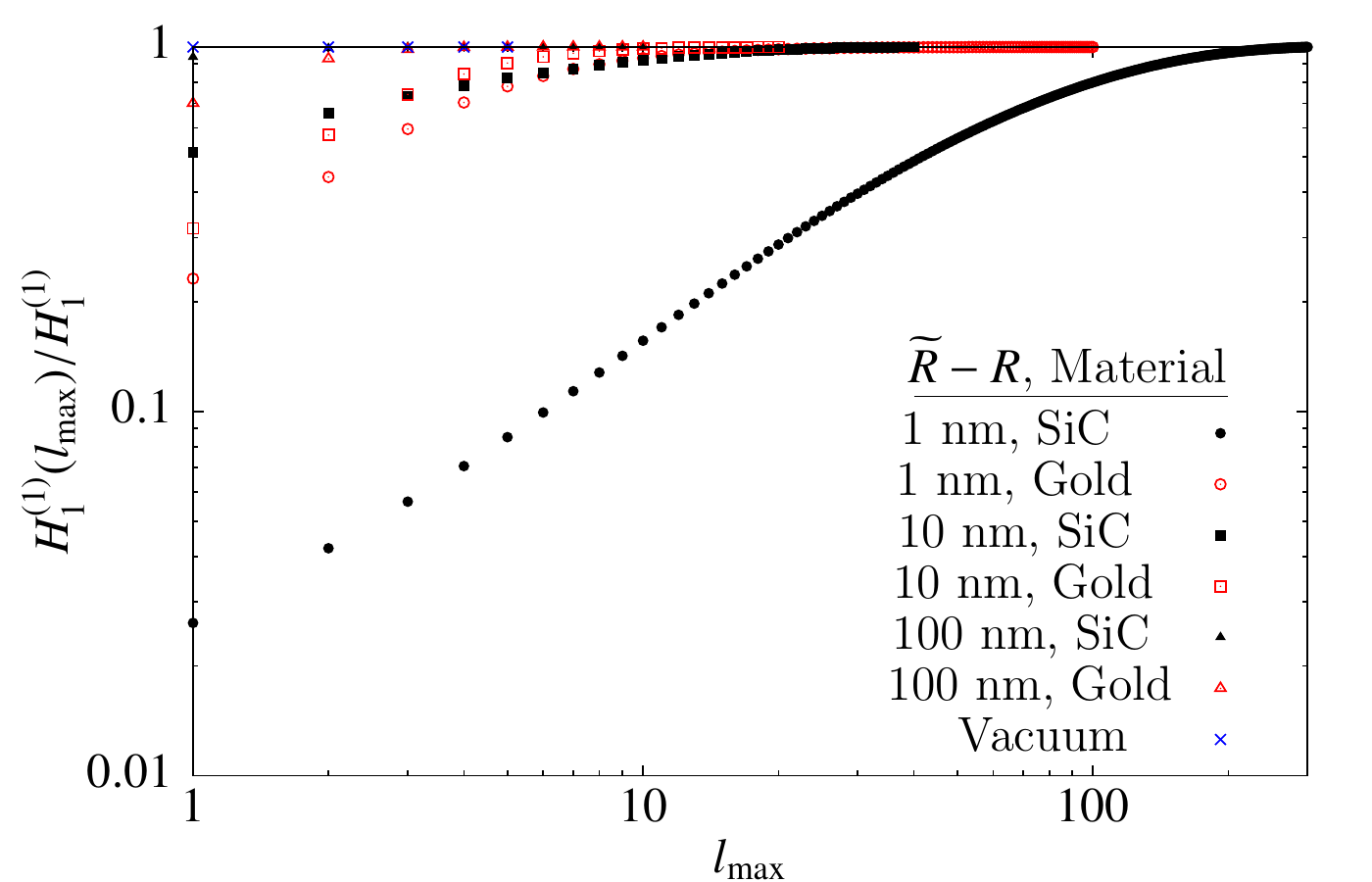}
\caption{\label{fig:CavityHR_gold_sph_conv}
Convergence of the heat radiation of a gold sphere with radius $ R = 10^{-1} \ \mu{\rm m} $ and temperature $ T_1 = 300 \ {\rm K} $ placed in the center of a spherical cavity with radius $ \widetilde{R} $ (see 
Fig.~\ref{fig:CavityHR_gold_sph}) as a function of the maximum multipole order used in the sum in Eq.~\eqref{eq:HR_sphere_cavity}, normalized by the exact value. For comparison, the corresponding curve 
for the radiation of the sphere in isolation is shown (labeled as \enquote{Vacuum}).}
\end{figure}
Figure~\ref{fig:CavityHR_gold_sph_conv} finally shows the convergence of exemplary points presented in Fig.~\ref{fig:CavityHR_gold_sph} with multipole order $l$. As expected from previous studies~\cite{Kruger2011, Narayanaswamy2008, Otey2011, Sasihithlu2011, Golyk2012}, the convergence slows down with decrease of $ \widetilde{R} - R $. 
For a SiC cavity, $ l_{\rm max}=300 $ is necessary for $ \widetilde{R} - R = 1 \ {\rm nm } $, while $l_{\rm max}=10$ suffices for $ \widetilde{R} - R = 100 \ {\rm nm } $. It is also remarkable that the 
convergence for a gold cavity is faster than for a SiC cavity. Lastly, we note that the results converge monotonically, which is in contrast to a nonmonotonic convergence of the HT between PPs in the presence of 
a sphere discussed in Ref.~\cite{Asheichyk2017}.  

We finish by providing simplified expressions for Eq.~\eqref{eq:HR_sphere_cavity}, based on previous literature~\cite{Kruger2011, Narayanaswamy2008, Otey2011, Sasihithlu2011, Golyk2012} as well as Fig.~\ref{fig:CavityHR_gold_sph_conv}. In the limit $ R \ll \lambda_{T_1} $ and $ R \ll \widetilde{R} $, we expect that the multiple reflections between sphere and cavity [the numerator in Eq.~\eqref{eq:HR_sphere_cavity}] can be neglected, and the sum in Eq.~\eqref{eq:HR_sphere_cavity} reduces to the term with $l=1$. Thus, in that limit, 
\begin{align}
\notag {H_1^{(1)}} = & -\frac{2\hbar}{\pi} \int_0^\infty d\omega \frac{\omega}{e^{\frac{\hbar\omega}{k_BT_1}}-1} \sum_{P=M,N}\\
&\times 3\left(\Re \widetilde{\mathcal{T}}_1^P + 1\right)\left(\Re \mathcal{T}_1^P + {\left|\mathcal{T}_1^P\right|}^2\right).
\label{eq:HR_sphere_cavity_limit}
\end{align}
If additionally the sphere is small compared to its skin depth (the point particle limit introduced above), we may further simplify, by using the  sphere's polarizability in Eq.~\eqref{eq:polarizability}, to obtain 
\begin{equation}
{H_{1pp}^{(1pp)}} = \frac{4\hbar}{\pi c^3} \int_0^\infty d\omega \frac{\omega^4}{e^{\frac{\hbar\omega}{k_BT_1}}-1} \Im (\alpha_1)\left[1 + \Re\widetilde{\mathcal{T}}_1^N\right].
\label{eq:HR_pp_sphere_cavity_final}
\end{equation}
Indeed, as regards Fig.~\ref{fig:CavityHR_SiC_sph}, the results of Eqs.~\eqref{eq:HR_pp_sphere_cavity_final} and \eqref{eq:HR_sphere_cavity} agree perfectly for $ \widetilde{R}- R > 1 \ \mu{\rm m} $. 
It is worth noting that the strong resonances seen in Fig.~\ref{fig:CavityHR_SiC_sph} can thus be computed by using the point particle approximation [Eq.~\eqref{eq:HR_pp_sphere_cavity_final}] of 
Eq.~\eqref{eq:HR_sphere_cavity}, which, \textit{a posteriori}, also justifies use of this approximation in Eq.~\eqref{eq:HT_PPs} to study the case depicted in Fig.~\ref{fig:config_plates}.

\section{Conclusion}
\label{sec:Conclusion}
In this paper, we studied heat radiation and transfer in confinement. In particular, we considered the heat transfer between two point particles placed between two parallel plates and the heat radiation of a 
sphere placed in the center of a spherical cavity. For both cases, we derived closed form expressions for the HT and HR and applied them to investigate several examples.

For the case of particles between  parallel plates,  the presence of the plates is found to enhance the HT dramatically (up to four orders of magnitude for SiC plates) 
and act as a wave guide for electromagnetic waves. While SiC plates show a strong, but a short (in distance between the particles) enhancement, gold plates provide a weaker, but a longer enhancement, showing a $ r^{-1} $ dependence for a large range of interparticle distance. The presence of plates is not additive in the sense that the results for two plates are distinct from the ones for a single plate studied in previous literature \cite{Dong2018,Messina2018}.

The emissivity of a gold sphere in a cavity can largely be understood in terms of the HT between two parallel surfaces, with the additional feature of a pronounced minimum if the cavity wall is made of gold as well.  For a SiC sphere placed in a cavity, we observe strong resonance behavior.  

Future work may study confinement in two dimensions. The simplest system that effectively removes two dimensions is a cavity made by an (infinitely long) cylinder. Further investigations can concern 
the HT between particles placed inside a spherical cavity. Also, other shapes, such as confinement by cubes or cones, may reveal interesting effects.

In summary, confined systems may open up new avenues for applications of heat radiation and transfer, in particular, with regard to the observed nonmonotonic effects.

\section*{Acknowledgments}
\label{sec:Acknowledgments}
We thank R.~L.~Jaffe and M.~Kardar for discussions and V.~A.~Golyk and M.~Kardar for initial discussions and results concerning the case of a sphere in a spherical cavity. This work was supported by MIT-Germany Seed Fund Grant No. 2746830 and Deutsche Forschungsgemeinschaft (DFG) Grant No. KR 3844/2-2. K.A. also acknowledges useful discussions with B.~M\"uller and H.~Soo, the financial support by the Physics Department of the University of Stuttgart and by International Max Planck Research School for Condensed Matter Science in Stuttgart, as well as the support by S.~Dietrich.


\begin{appendix}

\section{Green's function of two parallel plates}
\label{app:GF_plates}
We consider two identical semi-infinite parallel plates as depicted in Fig.~\ref{fig:config_plates} and aim to find the GF in the region between the plates. We work in Cartesian coordinate system with 
$ z $ axis perpendicular to the plates and place the plates such that the region between them is $ -d < z < 0 $. For this case, the initial expression for the GF, valid if both $ \vct{r}$ and $ \vct{r}' $ are in the region between the plates, reads as (note that there is an 
additional restriction, $ z > z' $, which we discuss later)~\cite{Johansson2011, Soo2016}
\begin{align}
\notag \mathbb{G}(\vct{r}, \vct{r}') = & \ \frac{i}{8\pi^2}\sum_P\int d^2k_{\perp}\frac{1}{k_z}\frac{1}{1-(F^P)^2e^{2ik_zd}}\\
\notag & \times \left[e^{i\vct{k}^+ \cdot \vct{r}}\vct{P}^+ + F^Pe^{i\vct{k}^-\cdot\vct{r}}\vct{P}^-\right]\\
& \otimes \left[e^{-i\vct{k}^+\cdot\vct{r'}}\vct{P}^+ + F^Pe^{2ik_zd}e^{-i\vct{k}^-\cdot\vct{r'}}\vct{P}^-\right],
\label{eq:GF2Plates1}
\end{align}
where $ P $ denotes polarization (magnetic $ M $ and electric $ N $), $ \vct{r}$ is the radius vector, $ d $ is the distance between the plates, $ \vct{k}^{\pm} = (\vct{k}_{\perp}, \pm k_z)^T $ is the 
wave vector in vacuum ($ k \equiv |\vct{k}^{\pm}| = \frac{\omega}{c} $), and $ k_z = \sqrt{k^2 - k_{\perp}^2} $. Symbol $ \otimes $ denotes the tensor product. Plane waves $ \vct{P}^{\pm} $ are defined 
as 
\begin{align}
& \vct{M}^+ = \vct{M}^- = \frac{1}{k_{\perp}}(-\hat{\vct{x}}k_y + \hat{\vct{y}}k_x),
\label{eq:vctM}\\
& \vct{N}^{\pm} = \frac{1}{k_{\perp}k}(\pm\hat{\vct{x}}k_xk_z \pm\hat{\vct{y}}k_yk_z - \hat{\vct{z}}k_{\perp}^2),
\label{eq:vctN}
\end{align}
where $ \hat{\vct{x}} $, $ \hat{\vct{y}} $, and $ \hat{\vct{z}} $ are spatial unit vectors in the respective directions. $ F^P $ are conventional Fresnel coefficients~\cite{Kruger2012, Muller2017, Asheichyk2017, Jackson1999}:
\begin{align}
& F^M = \frac{\sqrt{k^2-k_\perp^2}-\sqrt{\varepsilon k^2-k_\perp^2}}{\sqrt{k^2-k_\perp^2}+\sqrt{\varepsilon k^2-k_\perp^2}},
\label{eq:Fresnel_coeff_M}\\
& F^N = \frac{\varepsilon\sqrt{k^2-k_\perp^2}-\sqrt{\varepsilon k^2-k_\perp^2}}{\varepsilon\sqrt{k^2-k_\perp^2}+\sqrt{\varepsilon k^2-k_\perp^2}}.
\label{eq:Fresnel_coeff_N}
\end{align}

Green's function~\eqref{eq:GF2Plates1} is valid for the case $ z > z' $ only, which is a restriction for the free GF written in plane waves and hidden in expression~\eqref{eq:GF2Plates1}. To avoid this 
restriction, we can separate GF~\eqref{eq:GF2Plates1} into two parts, revealing the free GF $ \mathbb{G}_0(\vct{r}, \vct{r}') $ and the part due to the presence of the plates:
\begin{align}
\notag \mathbb{G}(\vct{r}, \vct{r}') = & \ \mathbb{G}_0(\vct{r}, \vct{r}') + \frac{i}{8\pi^2}\sum_P\int d^2k_{\perp}\frac{1}{k_z}\frac{1}{1-(F^P)^2e^{2ik_zd}}\\
\notag & \times \Big[F^Pe^{2ik_zd}e^{i\vct{k}^+\cdot\vct{r}}e^{-i\vct{k}^-\cdot\vct{r'}}\vct{P}^+\otimes\vct{P}^- \\
\notag & + F^Pe^{i\vct{k}^-\cdot\vct{r}}e^{-i\vct{k}^+\cdot\vct{r'}}\vct{P}^-\otimes\vct{P}^+\\
\notag & + (F^P)^2e^{2ik_zd}e^{i\vct{k}^-\cdot\vct{r}}e^{-i\vct{k}^-\cdot\vct{r'}}\vct{P}^-\otimes\vct{P}^-\\
& + (F^P)^2e^{2ik_zd}e^{i\vct{k}^+\cdot\vct{r}}e^{-i\vct{k}^+\cdot\vct{r}'}\vct{P}^+\otimes\vct{P}^+\Big].
\label{eq:GF2Plates2}
\end{align}
Since the free GF in Eq.~\eqref{eq:GF2Plates2} can be written in a closed form with no restrictions on $ \vct{r} $ and  $ \vct{r}' $ (see Eq.~(B1) in Ref.~\cite{Asheichyk2017}), GF~\eqref{eq:GF2Plates2} 
is valid for any $ \vct{r} $ and  $ \vct{r}' $ in the region between the plates.

Due to the fact that the system depicted in Fig.~\ref{fig:config_plates} is invariant under rotation around $ z $-axis, we can set, without loss of generality, the particles' positions to be 
$ \vct{r}_1 = (0, 0, -\frac{d}{2})^T $ and $ \vct{r}_2 = (r, 0, -\frac{d}{2})^T $, where $ r $ is the distance between the particles. Using these positions in GF~\eqref{eq:GF2Plates2} and 
performing the tensor products of plane waves~\eqref{eq:vctM} and~\eqref{eq:vctN}, we find 
\begin{align}
\notag \mathbb{G}&(\vct{r}_2, \vct{r}_1) = \mathbb{G}_0(\vct{r}_2, \vct{r}_1) + \frac{i}{4\pi^2}\int d^2k_{\perp}\frac{1}{k_z}\\
\notag & \times \Bigg\{\frac{F^Me^{ik_zd}}{1-F^Me^{ik_zd}}e^{ik_xr}\mathbb{M}_{\theta}\\
& \ \ \ \ + \frac{F^Ne^{ik_zd}}{1-(F^N)^2e^{2ik_zd}}e^{ik_xr}\left[\mathbb{N}_{\theta}'+F^Ne^{ik_zd}\mathbb{N}_{\theta}\right]\Bigg\},
\label{eq:GF2Plates3}
\end{align}
where 
\begin{align}
& \mathbb{M}_{\theta} = \frac{1}{k_{\perp}^2} 
\begin{pmatrix}
k_y^2 & -k_xk_y & 0\\
-k_xk_y & k_x^2 & 0\\
0 & 0 & 0
\end{pmatrix},
\label{eq:M_theta}\\
& \mathbb{N}_{\theta}' = \frac{1}{k^2k_{\perp}^2} 
\begin{pmatrix}
-k_x^2k_z^2 & -k_xk_yk_z^2 & 0\\
-k_xk_yk_z^2 & -k_y^2k_z^2 & 0\\
0 & 0 & k_{\perp}^4
\end{pmatrix},
\label{eq:N'_theta}\\
& \mathbb{N}_{\theta} = \frac{1}{k^2k_{\perp}^2} 
\begin{pmatrix}
k_x^2k_z^2 & k_xk_yk_z^2 & 0\\
k_xk_yk_z^2 & k_y^2k_z^2 & 0\\
0 & 0 & k_{\perp}^4
\end{pmatrix},
\label{eq:N_theta}
\end{align}
and subscript $ \theta $ denotes that the matrices depend on the polar angle $ \theta $ (in $ \vct{k}_{\perp} $ plane).

We now go to polar coordinates $ k_{\perp} $ and $ \theta $ in $ \vct{k}_{\perp} $ plane. First, we note that in Eq.~\eqref{eq:GF2Plates3} the angular dependence is only acquired by terms $ e^{ik_xr} $ 
and matrices $ \mathbb{M}_{\theta} $, $ \mathbb{N}_{\theta}' $, $ \mathbb{N}_{\theta} $. Second, the terms with an odd number of $ k_y $ in matrices 
$ \mathbb{M}_{\theta} $, $ \mathbb{N}_{\theta}' $, $ \mathbb{N}_{\theta} $ give zero after angular integration, because they produce odd functions in $ k_y $ in the total 
expression~\eqref{eq:GF2Plates3}. Therefore, the GF is diagonal, which is a consequence of the specific symmetry of the configuration chosen in Fig.~\ref{fig:config_plates}. Performing angular 
integration, we find
\begin{widetext}
\begin{align}
& \mathbb{M} \equiv \int_0^{2\pi}d\theta e^{ik_xr} \mathbb{M}_{\theta} = \frac{2\pi}{k_{\perp}r} 
\begin{pmatrix}
J_1(k_{\perp}r) & 0 & 0\\
0 & J_1(k_{\perp}r)-k_{\perp}rJ_2(k_{\perp}r) & 0\\
0 & 0 & 0
\end{pmatrix},
\label{eq:M}\\
& \mathbb{N}' \equiv \int_0^{2\pi}d\theta e^{ik_xr} \mathbb{N}_{\theta}' = \frac{2\pi}{k^2k_{\perp}r} 
\begin{pmatrix}
-k_z^2[J_1(k_{\perp}r)-k_{\perp}rJ_2(k_{\perp}r)] & 0 & 0\\
0 & -k_z^2J_1(k_{\perp}r) & 0\\
0 & 0 & k_{\perp}^3rJ_0(k_{\perp}r)
\end{pmatrix},
\label{eq:N'}\\
& \mathbb{N} \equiv \int_0^{2\pi}d\theta e^{ik_xr} \mathbb{N}_{\theta} = \frac{2\pi}{k^2k_{\perp}r} 
\begin{pmatrix}
k_z^2[J_1(k_{\perp}r)-k_{\perp}rJ_2(k_{\perp}r)] & 0 & 0\\
0 & k_z^2J_1(k_{\perp}r) & 0\\
0 & 0 & k_{\perp}^3rJ_0(k_{\perp}r)
\end{pmatrix},
\label{eq:N}
\end{align}
\end{widetext}
where $ J_i $ are Bessel functions of order $ i $.
Substituting Eqs.~\eqref{eq:M},~\eqref{eq:N'}, and~\eqref{eq:N} into Eq.~\eqref{eq:GF2Plates3}, we finally obtain for the GF
\begin{align}
\notag & \mathbb{G}(\vct{r}_2, \vct{r}_1) = \mathbb{G}_0(\vct{r}_2, \vct{r}_1) + \frac{i}{4\pi^2}\int_0^{\infty} dk_{\perp}\frac{k_{\perp}}{k_z}\\
& \times \Bigg\{\frac{F^Me^{ik_zd}}{1-F^Me^{ik_zd}}\mathbb{M} + \frac{F^Ne^{ik_zd}}{1-(F^N)^2e^{2ik_zd}}\left[\mathbb{N}'+F^Ne^{ik_zd}\mathbb{N}\right]\Bigg\}.
\label{eq:GF2PlatesFinal}
\end{align}
Note that the $ r $ dependence is in matrices $ \mathbb{M} $, $ \mathbb{N}' $, and $ \mathbb{N} $. Expression~\eqref{eq:GF2PlatesFinal} is used in formula~\eqref{eq:HT_PPs} to compute HT in 
Sec.~\ref{sec:HT_confinement_1D}. Integrals over $ k_{\perp} $ in GF~\eqref{eq:GF2PlatesFinal} and over $ \omega $ in formula~\eqref{eq:HT_PPs} are evaluated numerically.

\section{Derivation of the formula for the heat radiation of a sphere inside a spherical cavity}
\label{app:HR_sphere_cavity_derivation}
We derive formula~\eqref{eq:HR_sphere_cavity} from Eq. (44) in Ref.~\cite{Muller2017}. For definitions and expressions of scattering and translation matrices we refer the reader to 
Refs.~\cite{Rahi2009, Kruger2012, Muller2017, Asheichyk2017}. Naturally, we choose spherical basis for scattering and translation matrices. There are two important features of the derivation. First, 
since a sphere and a spherical cavity have the same origin, translation matrices are equal to the identity matrix. Second, the scattering matrices of both a sphere and a cavity are diagonal.

Starting from Eq.~(44) in Ref.~\cite{Muller2017} (note that we use minus of that equation, because we compute the heat emission, but not the heat absorption as done in 
Ref.~\cite{Muller2017}), we get
\begin{align}
\notag {H_1^{(1)}} = & \ -\frac{2\hbar}{\pi} \int_0^\infty d\omega \frac{\omega}{e^{\frac{\hbar\omega}{k_BT_1}}-1}\Re \Tr \Bigg\{\left[\widetilde{\mathcal{T}} + \mathcal{I}\right]\\
& \times \frac{1}{\mathcal{I}-\mathcal{T}\widetilde{\mathcal{T}}}\left[\frac{\mathcal{T}^{\dagger}+\mathcal{T}}{2}+\mathcal{T}\mathcal{T}^{\dagger}\right]\frac{1}{\mathcal{I}-\mathcal{T}^{\dagger}\widetilde{\mathcal{T}}^{\dagger}}\Bigg\},
\label{eq:HR_sphere_cavity_der_1}
\end{align}
where $ \mathcal{T} $ and $ \widetilde{\mathcal{T}} $ are scattering matrices of a sphere and a cavity, respectively, and $ \mathcal{I} $ is the identity matrix. Since all the matrices are diagonal, we have
\begin{align}
& \frac{\mathcal{T}^{\dagger}+\mathcal{T}}{2} = \frac{\mathcal{T}^*+\mathcal{T}}{2} = \Re \mathcal{T},
\label{eq:HR_sphere_cavity_der_2}\\
& \mathcal{T}\mathcal{T}^{\dagger} = \mathcal{T}\mathcal{T}^* = \left|\mathcal{T}\right|^2,
\label{eq:HR_sphere_cavity_der_3}\\
& \mathcal{T}^{\dagger}\widetilde{\mathcal{T}}^{\dagger} = \mathcal{T}^*\widetilde{\mathcal{T}}^*.
\label{eq:HR_sphere_cavity_der_4}
\end{align}
Moreover, the inverse matrices in Eq.~\eqref{eq:HR_sphere_cavity_der_1} are diagonal as well and can be thus rearranged with others.
Equation~\eqref{eq:HR_sphere_cavity_der_1} hence becomes
\begin{align}
\notag {H_1^{(1)}} = & \ -\frac{2\hbar}{\pi} \int_0^\infty d\omega \frac{\omega}{e^{\frac{\hbar\omega}{k_BT_1}}-1}\\
& \times \Tr \left\{\left[\Re \widetilde{\mathcal{T}} + \mathcal{I}\right] \left[\Re \mathcal{T}+\left|\mathcal{T}\right|^2\right] \frac{1}{\left|\mathcal{I}-\widetilde{\mathcal{T}}\mathcal{T}\right|^2}\right\}.
\label{eq:HR_sphere_cavity_der_5}
\end{align}
Next, we write the trace using matrix indexing:
\begin{align}
\notag & \Tr \left\{\left[\Re \widetilde{\mathcal{T}} + \mathcal{I}\right] \left[\Re \mathcal{T}+\left|\mathcal{T}\right|^2\right] \frac{1}{\left|\mathcal{I}-\widetilde{\mathcal{T}}\mathcal{T}\right|^2}\right\}\\
\notag & = \sum_{\mu}\left[\Re \widetilde{\mathcal{T}} + \mathcal{I}\right]_{\mu\mu} \left[\Re \mathcal{T}+\left|\mathcal{T}\right|^2\right]_{\mu\mu} \left(\frac{1}{\left|\mathcal{I}-\widetilde{\mathcal{T}}\mathcal{T}\right|^2}\right)_{\mu\mu}\\
\notag & = \sum_{\mu}\left[\Re \widetilde{\mathcal{T}}_{\mu\mu} + \mathcal{I}_{\mu\mu}\right] \left[\Re\mathcal{T}_{\mu\mu}+\left|\mathcal{T}_{\mu\mu}\right|^2\right]\\
& \hspace{1cm} \times \frac{1}{\left|\mathcal{I}_{\mu\mu}-\widetilde{\mathcal{T}}_{\mu\mu}\mathcal{T}_{\mu\mu}\right|^2},
\label{eq:HR_sphere_cavity_der_6}
\end{align}
where again we used diagonality of all the matrices. Since $ \mu = \{P, l, m\} $, 
where $ m = -l, -(l-1), \dots, 0, \dots, (l-1), l $, $ \mathcal{I}_{\mu\mu'} \equiv \delta_{\mu\mu'} = \delta_{PP'}\delta_{ll'}\delta_{mm'}$, 
and $ \mathcal{T}_{\mu\mu'} = \mathcal{T}_l^P\delta_{PP'}\delta_{ll'}\delta_{mm'} $, $ \widetilde{\mathcal{T}}_{\mu\mu'} = \widetilde{\mathcal{T}}_l^P\delta_{PP'}\delta_{ll'}\delta_{mm'} $, we have
\begin{align}
\notag & \Tr \left\{\left[\Re \widetilde{\mathcal{T}} + \mathcal{I}\right] \left[\Re \mathcal{T}+\left|\mathcal{T}\right|^2\right] \frac{1}{\left|\mathcal{I}-\widetilde{\mathcal{T}}\mathcal{T}\right|^2}\right\}\\
& = \sum_{P=M,N}\sum_{l=1}^{\infty}(2l+1) \left[\Re \widetilde{\mathcal{T}}_l^P + 1\right]\frac{\Re \mathcal{T}_l^P+\left|\mathcal{T}_l^P\right|^2}{\left|1-\widetilde{\mathcal{T}}_l^P\mathcal{T}_l^P\right|^2}.
\label{eq:HR_sphere_cavity_der_7}
\end{align}
Substituting Eq.~\eqref{eq:HR_sphere_cavity_der_7} into Eq.~\eqref{eq:HR_sphere_cavity_der_5}, we finally obtain formula~\eqref{eq:HR_sphere_cavity}.

\end{appendix}

\vspace{3.0cm}

\end{document}